 \documentclass[twocolumn,trackchanges]{aastex6}
\shorttitle{Siphon flow along an emerging magnetic flux tube}
\shortauthors{Requerey et al.}

\begin{document}


\title{Spectropolarimetric evidence for a siphon flow along an emerging magnetic flux tube}


\author{\textsc{
Iker~S.~Requerey$^{1,2,3}$, 
B.~Ruiz~Cobo$^{1,2}$, 
J.~C.~Del~Toro~Iniesta$^{3}$,
D.~Orozco~Su\'arez$^{3}$,
J.~Blanco~Rodr\'{\i}guez$^{4}$,
S.~K.~Solanki$^{5,6}$,
P.~Barthol$^{5}$,
A.~Gandorfer$^{5}$,
L.~Gizon$^{5,7}$,
J.~Hirzberger$^{5}$,
T.~L.~Riethm\"uller$^{5}$,
M.~van~Noort$^{5}$,
W.~Schmidt$^{8}$,
V.~Mart\'{\i}nez Pillet$^{9}$,
and M.~Kn\"olker$^{10}$
}}
\affil{
$^{1}$Instituto de Astrof\'{i}sica de Canarias, V\'{i}a L\'{a}ctea s/n, E-38205 La Laguna, Tenerife, Spain\\
$^{2}$Departamento de Astrof\'{i}sica, Universidad de La Laguna, E-38206 La Laguna, Tenerife, Spain\\
$^{3}$Instituto de Astrof\'{i}sica de Andaluc\'{i}a (CSIC), Apdo. de Correos 3004, E-18080 Granada, Spain\\
$^{4}$Grupo de Astronom\'{\i}a y Ciencias del Espacio, Universidad de Valencia, 46980 Paterna, Valencia, Spain\\
$^{5}$Max-Planck-Institut f\"ur Sonnensystemforschung, Justus-von-Liebig-Weg 3, 37077 G\"ottingen, Germany\\
$^{6}$School of Space Research, Kyung Hee University, Yongin, Gyeonggi, 446-701, Republic of Korea\\
$^{7}$Institut f\"ur Astrophysik, Georg-August-Universit\"at G\"ottingen, Friedrich-Hund-Platz 1, 37077 G\"ottingen, Germany\\
$^{8}$Kiepenheuer-Institut f\"ur Sonnenphysik, Sch\"oneckstr. 6, 79104 Freiburg, Germany\\
$^{9}$National Solar Observatory, 3665 Discovery Drive, Boulder, CO 80303, USA\\
$^{10}$High Altitude Observatory, National Center for Atmospheric Research,\footnote{The National Center for Atmospheric Research is sponsored by the National Science Foundation.} P.O. Box 3000, Boulder, CO 80307-3000, USA
}
\email{iker@iac.es}




\begin{abstract}

We study the dynamics and topology of an emerging magnetic flux concentration using high spatial resolution spectropolarimetric data acquired with the Imaging Magnetograph eXperiment on board the \textsc{Sunrise} balloon-borne solar observatory. We obtain the full vector magnetic field and the  line-of-sight (LOS) velocity through inversions of the Fe\,\textsc{i} line at 525.02\,nm with the SPINOR code. The derived vector magnetic field is used to trace magnetic field lines. Two  magnetic flux concentrations with different polarity and LOS velocities are found to be connected by a group of arch-shaped magnetic field lines. The positive polarity footpoint is weaker (1100\,G) and displays an upflow, while the negative polarity footpoint is stronger (2200\,G) and shows a downflow. This configuration is naturally interpreted as a siphon flow along an arched magnetic flux tube.

\end{abstract}




\section{Introduction}

Magnetic fields emerge on the solar surface in the form of arched magnetic flux tubes \citep{2014LRSP...11....3C}. If the arch exhibits a pressure difference at a given geometric height between the two footpoints a siphon flow can be driven along  the tube \citep{1968ZaMM...48..218M,1988ApJ...333..407T,1989ApJ...337..977M,1993ApJ...402..314M,1989A&A...222..297D,1991A&A...248..637D,1990ApJ...359..550T,1991ApJ...375..404T}.
The observational signature of a siphon flow in the solar photosphere is a pair of magnetic flux concentrations of opposite polarity with an upflow in one of them (the upstream footpoint) and a downflow along with a stronger magnetic field in the other (the downstream footpoint). Such a signature has been found by different authors \citep[e.g.,][]{1992A&A...261L..21R,1993A&A...279L..29D,2006ApJ...645..776U,2010AN....331..574B,2012A&A...537A.130B}. However, the direct measurement of the magnetic connectivity between the opposite footpoints has so far remained elusive. In the present paper we report on the three-dimensional magnetic topology of an emerging magnetic flux tube. The spectropolarimetric observations reveal a bunch of loop-shaped magnetic field lines that connect two regions of opposite polarities. The field strength and the LOS velocity within the footpoints are well described by the siphon-flow mechanism.

\section{Observations and data analysis}

\begin{figure*}
\includegraphics[width=\textwidth]{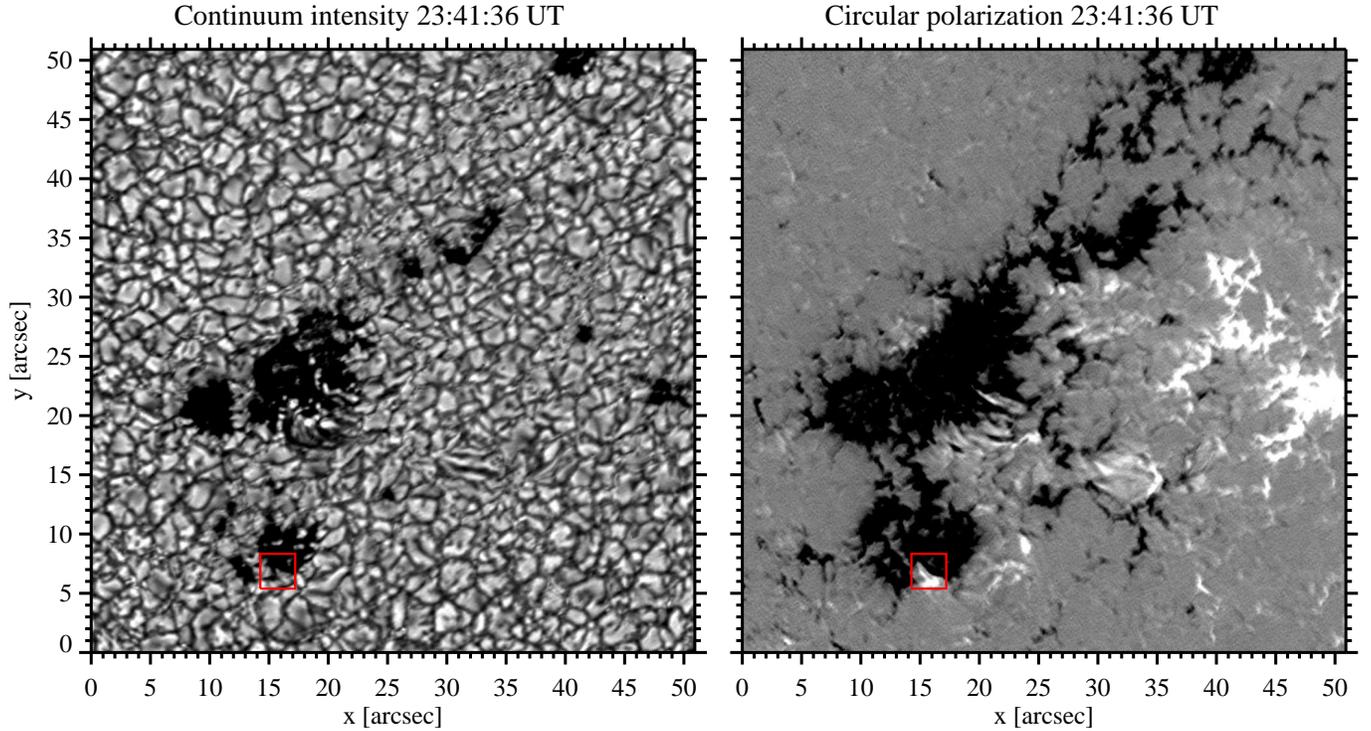}
\caption{Left panel: continuum intensity map, covering the FOV of IMaX of about 51\arcsec\ $\times$ 51\arcsec. Right panel: circular polarization map with a scale range of [-5,5]\,\% of the $I_{\rm c}$. The red solid rectangle, with a FOV of about 3\arcsec\ $\times$ 3\arcsec, illustrates the area studied throughout the paper.}
\label{fig1}
\end{figure*}

\begin{figure*}
\includegraphics[width=\textwidth]{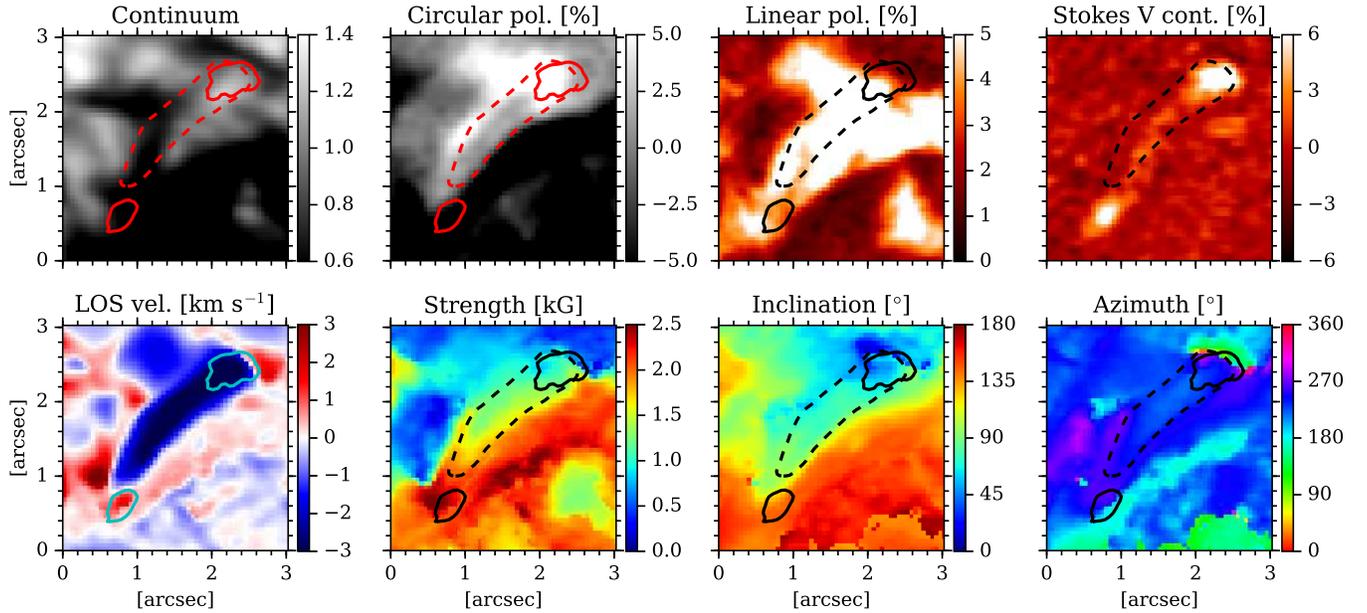}
\caption{Closeup of the red solid rectangle in Figure \ref{fig1}. From left to right and from top to bottom: continuum intensity, circular polarization, linear polarization, Stokes $V$ signal in the continuum, LOS velocity, strength, inclination and azimuth of the magnetic field in the local reference frame. Solid contours represent a continuum Stokes $V$ signal of 3\,\% of the $I_{\rm c}$. Dashed contours delineate regions containing upflow velocities stronger than  $1.75$\,km\,s$^{-1}$. The images has been flipped vertically with respect to Figure \ref{fig1}.}
\label{fig2}
\end{figure*}

We use high-quality spectropolarimetric data acquired with the Imaging Magnetograph eXperiment \citep[IMaX;][]{2011SoPh..268...57M}  during the second flight of the \textsc{Sunrise} balloon-borne solar observatory \citep{2010ApJ...723L.127S,2011SoPh..268....1B,2011SoPh..268..103B,2011SoPh..268...35G,2016ApJ.S}. IMaX is a dual-beam imaging spectropolarimeter with full Stokes vector capabilities in the Fe\,\textsc{i} line at 525.02\,nm  (Land\'{e} factor g\,=\,3). The line is sampled by a Fabry--P\'{e}rot interferometer at eight wavelength positions placed at $\lambda$\,=\,$-$12,\,$-$8,\,$-$4,\,0,\,$+$4,\,$+$8,\,$+$12, and $+$22.7\,pm from the line center, with 4 accumulations at each wavelength positions (V8-4 mode of IMaX). The polarization analysis is performed by two liquid crystal variable retarders and a polarizing beam splitter. The image sequences were recorded on 2013 June 12
between 23:39:10 and 23:55:37 UT (16.5\,minutes in length), with a cadence of 36.5\,s. The
full field of view (FOV) of the observations covers an area of 51\arcsec\ $\times$ 51\arcsec with a spatial sampling of 0\farcs 0545. The FOV was located within an emerging active region AR 11768 at a heliocentric angle of 21.56$\arcdeg$.

The science images were reconstructed using phase diversity measurements \citep{1982OptEn..21..829G,1996ApJ...466.1087P} as described by \citet{2011SoPh..268...57M}. After reconstruction, the spatial resolution has been estimated to be 0\farcs 15--0\farcs 18 and the noise level in each Stokes parameter is about 7\,$\times$\,10$^{-3}\,I_{\rm c}$ ($I_{\rm c}$ being the continuum intensity). Circular and linear polarization maps have been obtained by averaging the three blue and red wavelength points ($-$12,\,$-$8,\,$-$4,\,$+$4,\,$+$8,\,$+$12\,pm from the line center). For Stokes $V$ the three red wavelength points had their sign changed to avoid cancellation. We recover information of the vector magnetic field and LOS velocities through inversions of the full Stokes vector using the SPINOR code \citep{2000A&A...358.1109F}. A detailed description of the inversion procedure and LOS velocity calibration can be found in \citet{2016ApJ.S}. Height independent values for the three components of the magnetic field and LOS velocity are assumed. In addition, we derive the magnetic field inclination and azimuth values in the local reference frame using the \textsc{idl} routine \textit{r\_frame\_asp.pro} from the Advanced Stokes Polarimeter \citep{1992SPIE.1746...22E} \textsc{idl} library. Transformation to local coordinates depends on the azimuth. To resolve the 180$\arcdeg$ ambiguity we have minimized the angle between the observed field and a reference field direction, defined in our case by the direction of the line connecting the two footpoints \citep[see for instance, the Acute angle method in][]{2006SoPh..237..267M}.

\section{Results}
\label{sec3}

\begin{figure}[!b]
\includegraphics[scale=0.78]{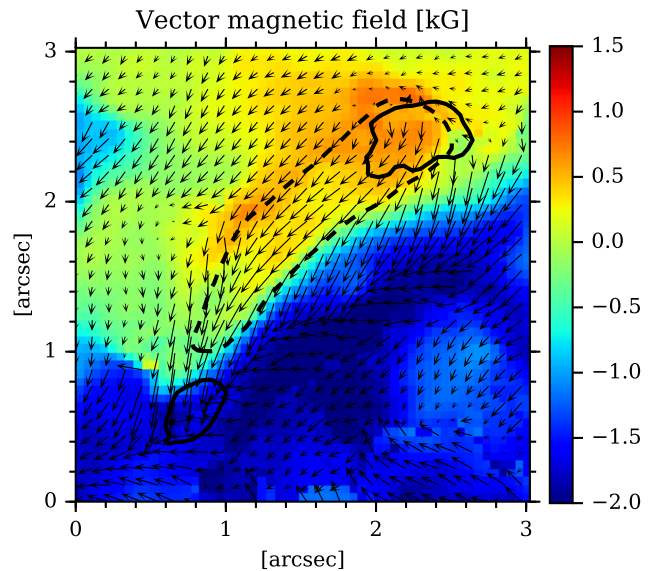}
\caption{Vector magnetic field in the local reference frame. The background image and the black arrows represent the vertical and horizontal component of the vector magnetic field, respectively.}
\label{fig3}
\end{figure}

\begin{figure*}
\includegraphics[width=\textwidth]{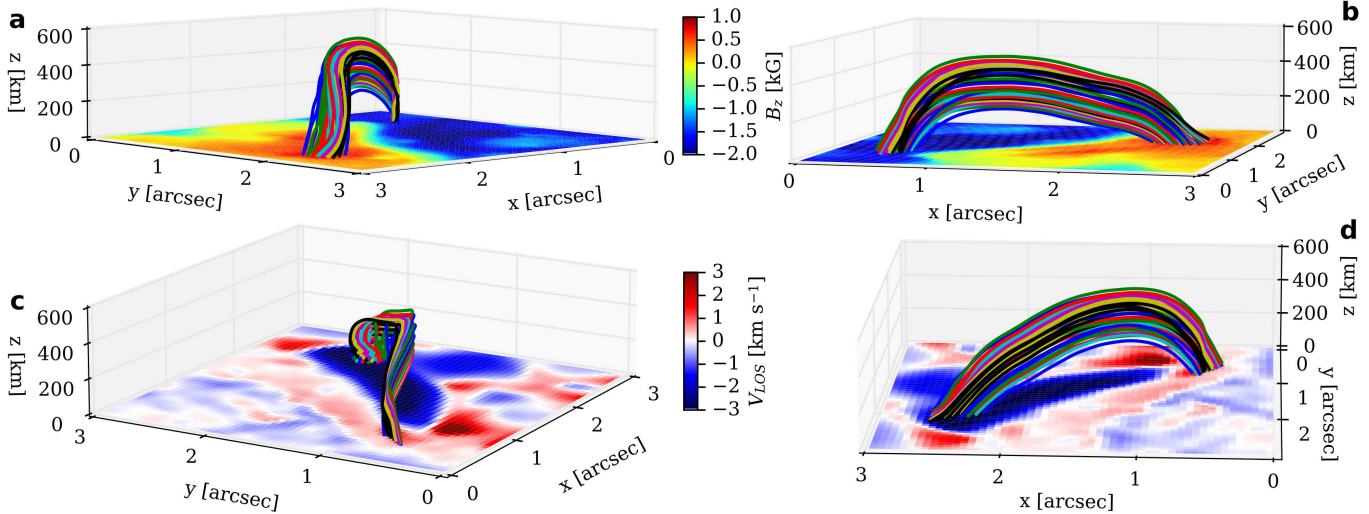}
\caption{Reconstructed magnetic field lines within the arched flux tube at four different view angles and $t=2.4$\,minutes. Colored images in the xy-planes show the vertical component of the vector magnetic field (\textbf{a}, \textbf{b}) and the LOS velocity (\textbf{c}, \textbf{d}). The color code of field lines has no meaning.}
\label{fig4}
\end{figure*}

\begin{figure}
\includegraphics[scale=0.65]{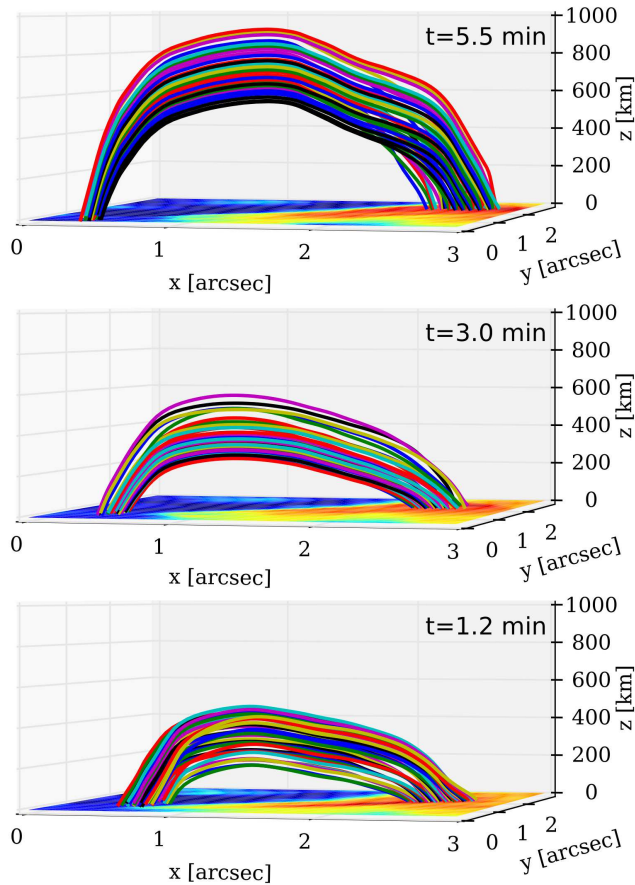}
\caption{Three snapshots of the evolution of the arched magnetic flux tube. Colored images in the xy-planes show the vertical component of the vector magnetic field. Time runs from bottom to top.}
\label{fig5}
\end{figure}

Figure \ref{fig1} displays the FOV observed by IMaX. The continuum intensity map (left panel) shows many different sized pores along with weaker and smaller magnetic features only visible in the circular polarization map (right panel). In the present paper we focus on a small region of 3\arcsec\ $\times$ 3\arcsec\ located within the red square in Figure \ref{fig1}. This area shows an elongated magnetic feature lying just outside the edge of an opposite polarity pore.

In Figure \ref{fig2} we zoom in on the region of interest. From left to right, the upper panels show the continuum intensity, $I_{\rm c}$, the circular and linear polarization maps, and the Stokes $V$ signal in the continuum, $V_{\rm c}$, all of them normalized to the mean quiet-Sun continuum intensity value. $V_{\rm c}$ displays two prominent  patches with signals above 3\,\% of the $I_{\rm c}$.  Such small-scale features were first detected in quiet-Sun areas by \citet{2010ApJ...723L.144B}. They interpreted these events as supersonic magnetic jets due to the large magnetized plasma velocities that are needed to explain them. Their physical origin has been ascribed to the process of magnetic reconnection by different authors \citet{2010ApJ...723L.144B,2013ApJ...768...69B,2011A&A...530A.111M,2013A&A...558A..30Q}. However, highly Doppler-shifted signals have also been observed at the footpoints of small-scale magnetic loops possibly associated with siphon flows \citet{2014A&A...569A..73Q}.  

The two $V_{\rm c}$ patches that we find in the upper right panel of Figure \ref{fig2} are delineated by solid contours in the other panels. As shown in the continuum intensity map, the uppermost one is located within a granule while the lowermost one lies at the edge of a pore. They appear related to opposite polarities in the circular polarization map and they are connected through an elongated strong linear polarization signal. Linear polarization flanked by two opposite circular polarization signals is reminiscent of magnetic loop-like features \citep{2007A&A...469L..39M,2007ApJ...666L.137C}. Thus, the strong $V_{\rm c}$ signals are most likely located at the two opposite polarity footpoints of an arched magnetic flux tube.

In the bottom panels of Figure \ref{fig2}  we show the physical parameters retrieved by the inversions. From left to right we display the LOS velocity, the magnetic field strength, the field inclination and the azimuth. The LOS velocity map displays a strong elongated upflow feature, which is outlined by dashed contours in the other panels. The rightmost end of this upflow is located within the positive polarity footpoint with a mean velocity of $-2.5$\,km\,s$^{-1}$. The upflow feature diagonally crosses the displayed zone until the LOS velocity map displays a localized downflow of 1\,km\,s$^{-1}$ at the negative polarity footpoint. The downflow reaches values as large as 5\,km\,s$^{-1}$ in the course of the time sequence. The upstream footpoint has a mean field strength of 1100\,G while the downstream footpoint displays a stronger field strength of 2200\,G.  Siphon flows along magnetic flux tubes are expected to show just such a signature: an upflow in one footpoint of the arch and a downflow and a stronger magnetic field in the other footpoint \citep{1991ApJ...375..404T}. We must bear in mind, however, that the siphon-flow hypothesis is only valid if the field strengths are measured at the same geometrical height. This is not the case here: the upstream footpoint is located in a bright granule while the downstream footpoint appears anchored in the edge of a dark pore. Accordingly, we can expect a sizeable height difference as the darkest footpoint probably has a Wilson depression of about a few hundred kilometers. Despite this, the large field strength difference (1100\,G) that we find between the two footpoints makes us suggest the siphon flow as the most plausible scenario.

\begin{figure*}
\includegraphics[width=\textwidth]{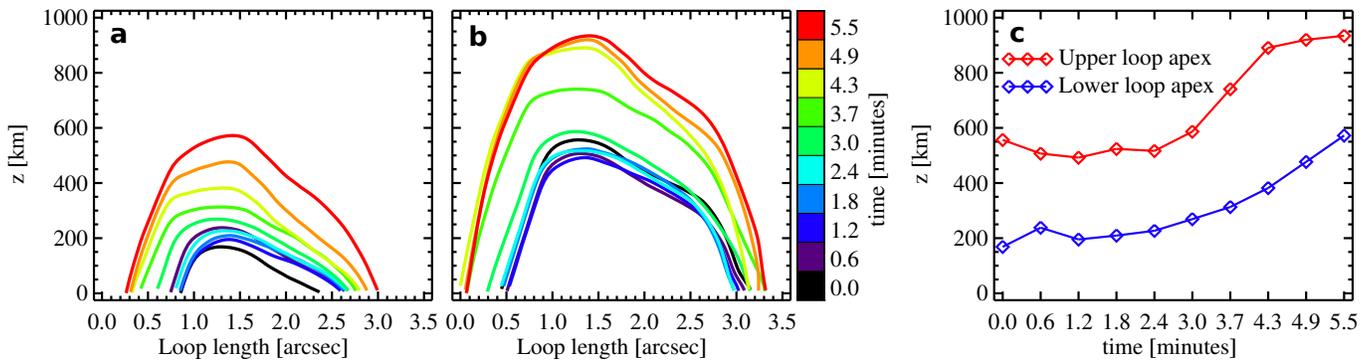}
\caption{Evolution of the arched magnetic flux tube in the solar atmosphere. The lowermost (\textbf{a}) and uppermost (\textbf{b}) magnetic field lines of the tube are represented with different colors at each time step.  The colors run from black to red as time increases. Temporal evolution of the apex height (\textbf{c}) for the lower and upper loops are shown by the blue and red curves, respectively.}
\label{fig6}
\end{figure*}

\subsection{Arched magnetic flux tube}
\label{sec31}

An important observational issue of siphon flows is to prove that the upstream and downstream footpoints are indeed connected by magnetic field lines. This is something that has not been demonstrated in previous works \citep{1992A&A...261L..21R,1993A&A...279L..29D,2006ApJ...645..776U,2010AN....331..574B,2012A&A...537A.130B}
Here, we take advantage of the high spatial resolution of IMaX data to circumvent this problem. 

In Figure \ref{fig3} we display the vector magnetic field in the local reference frame. The vertical and horizontal component of the vector magnetic field are shown by the background image and the black arrows, respectively. Positive and negative values in the background image imply that the field is directed  away from and into the solar surface. The vector magnetic field maps displays an arch-shaped structure along the elongated upflow feature (dashed contour). The magnetic field vector points up in the upstream footpoint, moves away almost horizontally (but still rising) towards the other end and points down in the downstream footpoint. 

Having the full vector magnetic field we can also reconstruct the three-dimensional structure of the arched magnetic flux tube in the solar atmosphere using a similar approach as in \citet{2003Natur.425..692S} and \citet{2010ApJ...714L..94M}. We trace magnetic field lines using the derived inclination and azimuth angles of the vector magnetic field. We only trace magnetic field lines that emanate from the positive footpoint. The footpoint is defined by a 7\,pixel ($\sim$\,280\,km) radius circle around the centroid of the $V_{\rm c}$ feature. Representative field lines are calculated starting from each pixel of the upstream footpoint. At this first position the geometrical reference height of the field lines is set to 0\,km. Then the nearest neighbor vector magnetic field is used to trace the lines.  Notice that this vector is assume to be constant with height as imposed by our inversions. The trajectories are integrated in space until they reach the same height at the other end of the arch. This procedure is independently repeated at each time step. The reconstructed field lines at $t=2.4$\,minutes are plotted in Figure \ref{fig4}. Colored images images in panels (\textbf{a}, \textbf{b}) and (\textbf{c}, \textbf{d}) display  the vertical component of the vector magnetic field and the LOS velocity, respectively. Both footpoints are connected by a bunch of field lines, which all together draw an arch-shaped magnetic loop. In Figure \ref{fig4}\,\textbf{d}, this upflow is observed beneath the field lines as they keep raising and it turns into a downflow when they bend and reach the other end of the tube. This suggests that matter is flowing all along the arched magnetic flux tube, from the weakest  footpoint towards the strongest one. However, a fraction of this upflow could also be due to an emerging process of the flux tube itself. 

\subsection{Temporal evolution}
\label{sec31}

The evolution of the reconstructed magnetic flux tube, as depicted by three snapshots in Figure \ref{fig5}, reveals that the magnetic feature is indeed rising through the solar atmosphere. At this point, a comment on the reliability of the reconstructed field lines is worth mentioning. The height-independent parameters retrieved by our inversions can be interpreted as averages of the  whole stratification weighted by the corresponding response functions \citep{1998ApJ...494..453W}. This means that, as the loop rises and abandons --at least partly--  the  formation region of the 525.02\,nm line, the uncertainty of the vector magnetic field that we ascribe to the uppermost part of the loop increases. After the first 3\,minutes, the field lines start to overpass the location of the downstream footpoint in the $V_{\rm c}$ map. This is an indication that beyond this time step the loop apex definitely lies above the formation region of the 525.02\,nm line, and hence the three-dimensional reconstruction of the magnetic field lines is not reliable anymore.

In order to better characterize this emergence process, in Figure \ref{fig6} we show the whole time evolution of the magnetic arch. For simplicity, we only display the lower (panel \textbf{a}) and upper (panel \textbf{b}) field lines of the whole bundle of lines. These are the ones with the lowest and highest apex heights, respectively. Each time step is represented by a different color as shown in the corresponding color bar. Temporal evolution of the apex height of both lower and upper field lines are also represented in panel \textbf{c} by the blue and red curves, respectively. The evolution of the lowermost field lines (panel \textbf{a}) and their apex height (blue curve in panel \textbf{c}) clearly shows that the flux tube is rising with time. In the most reliable part of the emergence phase, during the first 3\,minutes, the lowermost lines rise with a mean ascent velocity of 0.4\,km\,s$^{-1}$ and a footpoint separation speed of 1.5\,km\,s$^{-1}$. These values increase up to 1.1\,km\,s$^{-1}$ and 2.5\,km\,s$^{-1}$ when the first 5.5\,minutes are considered. This emergence process is not that steady for the uppermost field lines (panel \textbf{b} and red line in panel \textbf{c}). During the first 3\,minutes they display an apex height of about 500\,km, and then they suddenly rise to a larger value of $\sim$\,900\,km. This behavior suggests that we should not trust anything above 500\,km.

\section{Discussion and conclusions}

We have presented strong spectropolarimetric evidence for a siphon flow along an emerging magnetic flux tube. 
An elongated strong linear polarization signal is found between two regions of opposite polarity. An upflow is observed in one side and a downflow and a stronger magnetic field in the other. Similar observational signature of a siphon flow has been also reported by other authors in both the photosphere \citet{1992A&A...261L..21R,1993A&A...279L..29D} and the chromosphere \citep{2006ApJ...645..776U,2010AN....331..574B,2012A&A...537A.130B}. However, with our higher spatial resolution data, we have also been able to derive the full three-dimensional topology of the magnetic arch. This is, to our knowledge, the first time that the magnetic connectivity between the upstream and downstream footpoints of such a siphon flow has been directly measured. With a size of only a couple of arcseconds, our feature is one order of magnitude smaller than previously reported ones. This shows that the siphon flow effect also appears along small photospheric loops. In addition, our stable time series has enabled us to follow the rise of the loop for some 3 minutes until it leaves photospheric layers, thus reinforcing the fact that this effect is also observable in the chromosphere \citep{2006ApJ...645..776U,2010AN....331..574B,2012A&A...537A.130B}.

\acknowledgments

This work has been partially funded by the Spanish Ministerio de Econom\'{\i}a y Competitividad, 
through Projects No. ESP2013-47349-C6 and ESP2014-56169-C6, including a percentage from European FEDER funds.
The German contribution to \textsc{Sunrise} and its reflight was funded by the Max Planck Foundation, the Strategic Innovations Fund of the President of the Max Planck Society (MPG), DLR, and private donations by supporting members of the Max Planck Society, which is gratefully acknowledged. The National Solar Observatory (NSO) is operated by the Association of Universities for Research in Astronomy (AURA) Inc. under a cooperative agreement with the National Science Foundation.
The HAO contribution was partly funded through NASA grant number NNX13AE95G. This work was partly supported by the BK21 plus program through the National Research Foundation (NRF) funded by the Ministry of Education of Korea.
\listofchanges

\end{document}